\documentclass[10pt]{article}

\usepackage[utf8]{inputenc}
\usepackage[english]{babel}
\usepackage{graphicx,amsmath,amsfonts,amssymb,dcolumn,amsthm,slashed,bbm,xspace,pgffor,tikz,mathbbol,latexsym,cite,braket}
\usepackage{microtype}
\usepackage{lmodern}
\usepackage[normalem]{ulem}
\usepackage{xcolor}
\usepackage[colorlinks=true,citecolor=blue,urlcolor=blue,linkcolor=blue]{hyperref}
\usepackage[hang]{footmisc} %removes the footer margin space
\usepackage[affil-it]{authblk}
\usepackage[compat=1.1.0]{tikz-feynhand}
\usepackage{multicol}

\numberwithin{equation}{section}

\begin{document}

\title{\textbf{Infinitesimal Gribov copies in gauge-fixed topological Yang-Mills theories}}

\author{D.~Dudal$^{a, b}$\thanks{\href{mailto:david.dudal@kuleuven.be}{david.dudal@kuleuven.be}}, C.~P.~Felix$^{a, c}$\thanks{\href{mailto:caroline.felix@kuleuven.be}{caroline.felix@kuleuven.be}}, O.~C.~Junqueira$^{a,d}$\thanks{\href{mailto:octavio@if.uff.br}{octavio@if.uff.br}}, D.~S.~Montes$^d$\thanks{\href{mailto:douglasmontes@id.uff.br}{douglasmontes@id.uff.br}}, A.~D.~Pereira$^{d,e}$\thanks{\href{mailto:a.pereira@thphys.uni-heidelberg.de}{adpjunior@id.uff.br}}, G.~Sadovski$^{d,f}$\thanks{\href{mailto:guilherme.sadovski@oist.jp}{guilherme.sadovski@oist.jp}}, R.~F.~Sobreiro$^d$\thanks{\href{mailto:rodrigo\_sobreiro@id.uff.br}{rodrigo\_sobreiro@id.uff.br}}, A.~A.~Tomaz$^g$\thanks{\href{mailto:anderson.tomaz@ufabc.edu.br}{anderson.tomaz@ufabc.edu.br}} }
\affil{\footnotesize $^{a}$ KU Leuven Campus Kortrijk---Kulak, Department of Physics, Etienne Sabbelaan 53 bus 7657, 8500 Kortrijk, Belgium\\
 $^{b}$ Ghent University, Department of Physics and Astronomy, Krijgslaan 281-S9, 9000 Gent, Belgium\\
 $^{c}$ Chung Yuan Christian University, No. 200, Zhongbei Road, Zhongli District, Taoyuan City, 320, Taiwan\\
 $^{d}$ UFF---Universidade Federal Fluminense, Instituto de F\'isica, Av.~Litoranea s/n, 24210-346, Niter\'oi, RJ, Brasil\\
 $^{e}$ Institute for Theoretical Physics, University of Heidelberg, Philosophenweg 12, 69120 Heidelberg, Germany\\
 $^{f}$ OIST---Okinawa Institute of Science and Technology, 1919-1 Tancha, Onna-son, Kunigami-gun, Okinawa 904-0495, Japan \\
 $^{g}$ Center for Natural and Human Sciences---Federal University of ABC, Av. dos Estados 5001, Santo Andr\'e, S\~ao Paulo, SP, Brasil}

\date{}
\maketitle

\begin{abstract}
We study the Gribov problem in four-dimensional topological Yang-Mills theories following the Baulieu-Singer approach in the (anti-)self-dual Landau gauges. This is a gauge-fixed approach that allows to recover the topological spectrum, as first constructed by Witten, by means of an equivariant (or constrained) BRST cohomology.
As standard gauge-fixed Yang-Mills theories suffer from the gauge copy (Gribov) ambiguity, one might wonder if and how this has repercussions for this analysis. The resolution of the small (infinitesimal) gauge copies, in general, affects the dynamics of the underlying theory. In particular, treating the Gribov problem for the standard Landau gauge condition in non-topological Yang-Mills theories strongly affects the dynamics of the theory in the infrared. In the current paper, although the theory is investigated with the same gauge condition, the effects of the copies turn out to be completely different. In other words: in both cases, the copies are there, but the effects are very different.  As suggested by the tree-level exactness of the topological model in this gauge choice, the Gribov copies are shown to be inoffensive at the quantum level. To be more precise, following Gribov, we discuss the path integral restriction to the Gribov horizon. The associated gap equation, which fixes the so-called Gribov parameter, is however shown to only possess a trivial solution, making the restriction obsolete. We relate this to the absence of radiative corrections in both gauge and ghost sectors. We give further evidence by employing the renormalization group which shows that, for this kind of topological model, the gap equation indeed forbids the introduction of a massive Gribov parameter.
\end{abstract}

\section{Topological Yang-Mills theories in the (anti-)self-dual Landau gauges}

During the early eighties, Donaldson constructed a whole new class of topological invariants as integrals of differential forms over the moduli space of instantons \cite{Donaldson:1983wm}. The Donaldson polynomials are of the utmost importance in the classification of four-manifolds as they keep track of the smoothly inequivalent ways one may cover a topological space with local charts. This created a new toolbox to study the so-called ``exotic'' manifolds \cite{Maluga:2007}, a.k.a.~manifolds with non-standard differential (smooth) structures. The simplest examples are the ``exotic'' $\mathbb{R}^4$'s: four-manifolds homeomorphic to the usual Euclidean space, $\mathbb{R}^4$, but not diffeomorphic to it.
		
		The classification of four-manifolds is not only an abstract topic reserved for mathematicians. The physics on exotic manifolds has also being investigated with results ranging from particle physics to cosmology, \cite{Brans:1992mj,Brans:1994hq,Maluga:2007,Asselmeyer-Maluga:2018ywa,Sladkowski:2009dc,Asselmeyer-Maluga:2017tbn}. Moreover topology-changing processes might play a relevant role in quantum gravity and QCD, to name only these two examples. For instance, the knowledge of topologically inequivalent four-manifolds might be fundamental to define the physically inequivalent states in some quantum gravity models \cite{Duston:2009ma,AsselmeyerMaluga:2010ti,Asselmeyer-Maluga:2016foc}. On the other hand, the moduli space of instantons represents a huge degeneracy of the QCD vacuum.  Topology-changing processes among these vacua, a famous non-perturbative effect, can explain the anomalous $U(1)$ axial symmetry \cite{tHooft:1986ooh,tHooft:1976rip} and is related to the strong CP problem.

In what follows, we will consider a $SU(N)$ topological Yang-Mills theory over a four-dimensional Euclidean spacetime. This theory was first put forward by Witten as an exact local quantum field theory description of the Donaldson polynomials \cite{Witten:1988ze}.
The observables of such a theory are not dynamically propagating field excitations (there are none), rather they are exclusively given by topologically invariant Donaldson polynomials \cite{Donaldson:1983wm}. A nice review paper on general aspects is \cite{vanBaal:1989aw}.

Here, we will not be directly concerned with the Witten construction itself, but rather with the local BRST description of such topological quantum field theories, as it was introduced in \cite{Baulieu:1988xs} and further worked out in \cite{Ouvry:1988mm,Brandhuber:1994uf}. The goal of such program is to characterize the topological degrees of freedom using local quantum (gauge) field theory, that is, via an action that is constrained via various Ward identities. The (topological) observables of the theory are then defined as the elements of a BRST cohomology (gauge invariant operators) that also do not depend on the ghost field $c$, see \cite{vanBaal:1989aw,Ouvry:1988mm}, thereby defining an equivariant cohomology. That the standard cohomology is trivial can be seen by using shifted variables \cite{Ouvry:1988mm} in which case the new variables always appear as doublets, i.e., cohomologically trivial \cite{Piguet:1995er}. In \cite{Delduc1996}, see also \cite{Boldo:2003jq}, it was shown one can evade the complications of having to deal with expliciting an equivariant cohomology, rather the observables in field space can also be defined from a constrained BRST cohomological analysis. The particular space of observables was shown to contain only polynomials of the field $\phi$ (see later) and their correspondent descent equations (see later), without spacetime derivatives. These correspond precisely to the Donaldson polynomials upon explicit further evaluation, as it can be found in e.g.~\cite{Weis:1997kj}. As such, the Witten construction is recovered in this alternative language, providing a quantum field theory description of the Donaldson invariants.

Once the topological Yang-Mills theory is reformulated in the BRST language, one automatically encounters the issue of gauge fixing. For the usual (non-topological) non-Abelian Yang-Mills theories, the gauge fixing (Baulieu-Singer) procedure is hampered by the Gribov obstruction: in any covariant gauge, there are multiple solutions to the gauge condition \cite{Gribov:1977wm,Singer:1978dk}. A procedure to remove the particular subclass of the small gauge copies was proposed first by Gribov at the semi-classical level, and later on improved upon by Zwanziger in a series of papers, see e.g.~\cite{DellAntonio:1989wae,DellAntonio:1991mms,Zwanziger:1989mf,Zwanziger:1992qr} or \cite{Vandersickel:2012tz} for a review. In brief, one restricts the domain of integration for the gauge connection to a smaller region, encapsulated by the region of all gauge connections in the chosen specific gauge. Modulo some mild assumptions, this restriction can be implemented at the level of the action, leading to the Gribov-Zwanziger (GZ) quantization scheme. Its main effect is the introduction of a new, massive, Gribov parameter which is dynamically fixed via a suitable gap equation. Obviously, the introduction of a massive parameter in the originally massless gauge theory strongly influences the infrared dynamics, with potential ramifications to understand typical non-perturbative phenomena such as (de)confinement, etc. We refer the interested reader to the vast literature on these topics. 

A major drawback of the original GZ approach was the loss of BRST invariance. Though, more recently, some of us were involved in developing a BRST invariant formulation of the GZ procedure, see for instance \cite{Capri:2015ixa,Capri:2016aqq}. Doing so, the Gribov mass gained a BRST invariant status and as such it explicitly enters the BRST cohomology of physical operators.

If such massive Gribov parameter would also be there in the topological case, it could influence the constrained cohomology construction and, potentially, invalidate the algebraic identification of the Donaldson polynomials from a BRST perspective, i.e., the spectrum might change and be no longer of a topological nature. The main result of our paper will be to show that, precisely because of the topological nature of the action, the dynamical resolution of the Gribov problem becomes trivial at the end of the day, a fact made clear by a dynamically enforced vanishing of the Gribov parameter. In such case, the topological version of the GZ action reduces to the original one and we are back at the Baulieu-Singer-et al construction.

\subsection{The BRST symmetry and gauge fixing}
The manifold we construct the theory on is a four-dimensional spacetime which is assumed to be Euclidean and flat. Following the Baulieu-Singer approach \cite{Baulieu:1988xs}, the topological action $S_0[A]$  in four-dimensional space-time\footnote{$S_0[A]$ is the Pontryagin action, $S_0[A] = 32 \pi^2 n$, in which $n$ is the winding number that labels topologically inequivalent field configurations \cite{Coleman:1978ae}.},
\begin{equation} \label{Pontryagin}
S_0[A] = \frac{1}{2}\int d^4x\, F^a_{\mu\nu} \widetilde{F}^a_{\mu\nu}\;,
\end{equation}
where $A \equiv A^a_\mu$ is the gauge field, $F^a_{\mu\nu} = \partial_\mu A^a_\nu - \partial_\nu A^a_\mu + gf^{abc} A^b_\mu A^c_\nu$ is the field strength, with $\widetilde{F}^a_{\mu\nu} = \frac{1}{2} \varepsilon_{\mu\nu\alpha\beta} F^a_{\alpha\beta}$ being its dual,  has a reducible gauge symmetry to be fixed, these are:

(i) the gauge field symmetry,
        \begin{equation}
            \delta A_{\mu}^{a}=D_{\mu}^{ab}\omega^{b}+\alpha_{\mu}^{a}\;;
            \label{eqn:gluon-symmetry}
        \end{equation}

(ii) the topological parameter symmetry,
        \begin{equation}
            \delta\alpha_{\mu}^{a}=D_{\mu}^{ab}\lambda^{b}\;;
            \label{eqn:top-parameter-symmetry}
        \end{equation}
For later usage, let us notice that, a fortiori, the field strength itself transforms as\footnote{The antisymmetrization index notation here employed means that, for a generic tensor, $S_{[\mu\nu]}=S_{\mu\nu}-S_{\nu\mu}$.}
        \begin{equation}
            \delta F_{\mu\nu}^{a}=-gf^{abc}\omega^{b}F_{\mu\nu}^{c}+D_{[\mu}^{ab}\alpha_{\nu]}^{b}\;;
            \label{eqn:F-symmetry}
        \end{equation}
where $D_\mu^{ab} \equiv \delta^{ab}\partial_\mu - gf^{abc}A^c_\mu$ is the covariant derivative in the adjoint representation of the Lie group $G$, $g$ is the coupling constant, $f^{abc}$ are the structure constants of the gauge group and $\omega^a$, $\alpha^a_\mu$ and $\lambda^a$ are the infinitesimal $G$-valued gauge parameters. The first parameter reflects the usual Yang-Mills symmetry of $S[A]$, whereas the second one is associated to the fact that $S[A]$ is a topological invariant.

Said otherwise, the transformation \eqref{eqn:gluon-symmetry} actually consists of 2 parts: the standard local gauge symmetry variation and a ``topological shift'', expressing that the theory is essentially invariant under arbitrary variations of the gauge field, see also \cite{Ouvry:1988mm}, where this shift was directly expressed in terms of a Grassmann quantity $\psi_\mu$. As we wish to preserve the bosonic nature of the classical symmetries, we may write for this shift $\delta_{shift} A_\mu= \varepsilon \psi_\mu\equiv \alpha_\mu$ with $\varepsilon$ a Grassmann constant. 

The invariances \eqref{eqn:gluon-symmetry}-\eqref{eqn:top-parameter-symmetry} constitute a typical example of a reducible gauge symmetry, which BRST quantization requires a certain care, in particular the introduction of ``ghosts of ghosts'', see \cite{Henneaux:1992ig,Piguet:1995er}. Given the freedom in $\omega$, $\alpha_\mu$ and $\lambda$, we need 3 sets of gauge conditions, to be specified soon hereafter. Following the BRST quantization procedure, the gauge parameters present in the gauge transformations \eqref{eqn:gluon-symmetry}-\eqref{eqn:top-parameter-symmetry} are promoted to ghost fields: $\omega^a \rightarrow c^a$, $\alpha^a_\mu \rightarrow \psi^a_\mu$, and $\lambda^a \rightarrow \phi^a$; $c^a$ is the well-known Faddeev-Popov (FP) ghost; $\psi^a_\mu$ is a topological fermionic ghost; and $\phi^a$ is a bosonic ghost. The corresponding BRST transformations are
\begin{eqnarray}
sA_\mu^a&=&-D_\mu^{ab}c^b+\psi^a_\mu,\nonumber\\
sc^a&=&\frac{g}{2}f^{abc}c^bc^c+\phi^a,\nonumber\\
s\psi_\mu^a&=&gf^{abc}c^b\psi^c_\mu+D_\mu^{ab}\phi^b,\nonumber\\
s\phi^a&=&gf^{abc}c^b\phi^c,\label{brst1}
\end{eqnarray}
from which one can easily check the nilpotency of the BRST operator, i.e., $s^2 = 0$, thus defining the cohomology of the theory.

In order to fix the gauge symmetries we define the following  set of BRST doublets:
\begin{eqnarray}
s\bar{c}^a&=&b^a\;,\;\;\;\;\;\;\;\;sb^a\;=\;0\;,\nonumber\\
s\bar{\chi}^a_{\mu\nu}&=&B_{\mu\nu}^a\;,\;\;sB_{\mu\nu}^a\;=\;0\;,\nonumber\\
s\bar{\phi}^a&=&\bar{\eta}^a\;,\;\;\;\;\;\;\;s\bar{\eta}^a\;=\;0\;,\label{brst2}
\end{eqnarray}
where $\bar{\chi}^a_{\mu\nu}$ and $B_{\mu\nu}^a$ are (anti-)self-dual fields according to the (negative) positive sign in \eqref{gf3}. Working in the (anti-)self-dual Landau gauges (ASDL) amounts to considering the constraints \cite{Brandhuber:1994uf}
\begin{eqnarray}
    \label{eqn:gauge-fixings}
    \partial_{\mu} A_{\mu}^{a} &=& 0\;, \label{gf1}\\
    \partial_{\mu} \psi_{\mu}^{a} &=& 0\;,\label{gf2} \\
    F_{\mu\nu}^{a} \pm \widetilde{F}_{\mu\nu}^{a} &=& 0\;.\label{gf3}
\end{eqnarray}
Beyond the gauge fixing of the topological ghost \eqref{gf2}, we must interpret the requirement of two extra gauge fixings due to the fact that the gauge field possesses two independent gauge symmetries. In this sense the condition \eqref{gf1} fixes the usual Yang-Mills symmetry $\delta A^a_\mu = D_\mu^{ab}\omega^b$, and the second one, \eqref{gf3}, the topological shift $\delta A^a_\mu = \alpha^a_\mu$. The (anti-)self-dual condition for the field strength is convenient to identify the well-known observables of topological theories in four dimensions (see \cite{vanBaal:1989aw}) known as Donaldson polynomials \cite{Donaldson:1983wm}, that are described in terms of the instantons --- in which we are interested in here. This condition on $F_{\mu\nu}$ \eqref{gf3}, which is indirectly a condition on the gauge field as $F_{\mu\nu}$  only depends on $A^ a_\mu$, corresponds to the gauge fixing of the field strength itself, because $F^a_{\mu\nu}$ also transforms as a gauge field, cf.~\eqref{eqn:F-symmetry}.  The first gauge condition on $A^a_\mu$ fixes the information about its divergence while the second one restricts its curl freedom, in such a way that, from the point of view of the four-dimensional Helmholtz theorem \cite{Woodside:1999yj}, the gauge field is well-defined---disregarding the Gribov copies for a moment.

The complete gauge-fixed action in the (A)SDL gauges then takes the form
\begin{equation}\label{Stotal}
S[\Phi] = S_0[A] + S_{gf}[\Phi]\;,
\end{equation}
for all fields $\Phi \equiv \{A, \psi, c, \phi, \bar{c}, b, \bar{\phi}, \bar{\eta}, \bar{\chi}, B\}$, where
\begin{eqnarray}\label{S_gf}
S_{gf}\left[\Phi\right]&=&s\int d^4z\left[\bar{c}^a\partial_\mu A_\mu^a+\frac{1}{2}\bar{\chi}^a_{\mu\nu}\left(F_{\mu\nu}^a\pm\widetilde{F}_{\mu\nu}^a\right)+\bar{\phi}^a\partial_\mu\psi^a_\mu\right]\nonumber\\
&=&\int d^4z\left[b^a\partial_\mu A_\mu^a+\frac{1}{2}B^a_{\mu\nu}\left(F_{\mu\nu}^a\pm\widetilde{F}_{\mu\nu}^a\right)+
\left(\bar{\eta}^a-\bar{c}^a\right)\partial_\mu\psi^a_\mu+\bar{c}^a\partial_\mu D_\mu^{ab}c^b+\right.\nonumber\\
&-&\left.\frac{1}{2}gf^{abc}\bar{\chi}^a_{\mu\nu}c^b\left(F_{\mu\nu}^c\pm\widetilde{F}_{\mu\nu}^c\right)-\bar{\chi}^a_{\mu\nu}\left(\delta_{\mu\alpha}\delta_{\nu\beta}\pm\frac{1}{2}\epsilon_{\mu\nu\alpha\beta}\right)D_\alpha^{ab}\psi_\beta^b+\bar{\phi}^a\partial_\mu D_\mu^{ab}\phi^b+\right.\nonumber\\
&+&\left.gf^{abc}\bar{\phi}^a\partial_\mu\left(c^b\psi^c_\mu\right)+\frac{\alpha}{2}b^a b^a+\frac{\beta}{4} B_{\mu\nu}^a B_{\mu\nu}^a\right]\;.\label{gfaction}
\end{eqnarray}
The last two terms, which are BRST trivial, were added for later computational convenience; it is understood that, at the end, the limits $\beta\to0$, $\alpha\to0$ must be taken. We relied on the standard BRST quantization lore here \cite{Baulieu:1983tg,Piguet:1995er}, but it can be easily checked that upon integration over the various multipliers/auxiliary fields, the gauge fixing conditions are retrieved under the form of appropriate $\delta$-functions and corresponding Jacobians\footnote{See Appendix B for some details.}, representing the ``unities'' of the textbook Faddeev-Popov quantization procedure, at least for the here considered Landau gauge. The BRST method is however more convenient and more general than the Faddeev-Popov procedure. Indeed, not every gauge fixing needs to be of the ``unity type'', a famous example being the non-linear gauges of the Baulieu--Thierry-Mieg type \cite{Baulieu:1981sb}. 

The action \eqref{gfaction} enjoys a rich set of Ward identities, including the vector supersymmetry\footnote{It should be understood that the vector supersymmetry is present for $\alpha=\beta=0$.} \cite{Brandhuber:1994uf}. One can also prove that the gauge field propagator vanishes to all orders in perturbation theory \cite{Junqueira:2017zea}, and, consequently, that the theory is completely free of radiative corrections. In other words, that the theory is tree-level exact \cite{Junqueira:2018xgl}.

From the action \eqref{gfaction}, the $\bar{\chi}$ equation of motion gives
\begin{equation} \label{zeromodes1}
\Theta^{ab}_{\pm, \mu\nu\beta} \psi^b_\beta = 0\;,
\end{equation}
with
\begin{equation}\label{thetadef}
    \Theta^{ab}_{\pm, \mu\nu\beta} = (\delta_{\mu\alpha} \delta_{\nu\beta} - \delta_{\nu\alpha}\delta_{\mu\beta} \pm \epsilon_{\mu\nu\alpha\beta}) D^{ab}_{\alpha}\;,
\end{equation}
while the $\bar{\eta}$ equation of motion gives
\begin{equation}\label{zeromodes2}
\partial_\mu \psi^a_\mu = 0\;.
\end{equation}
The two last equations are precisely the two equations concerning the infinitesimal instanton moduli. We obtain here the same situation as present in the Witten version of the theory, see \cite{Witten:1988ze}; the only difference is the gauge choice. If we want to reproduce exactly the Witten equations, we should use the gauge constraint $D^{ab}_\mu \psi^a_\mu = 0$, instead of the Landau one.  As this is a gauge condition anyhow, physics should not depend on it. The reason to prefer the Landau gauge is the associated larger symmetry content, in particular the vector supersymmetry, as it was originally noticed in \cite{Brandhuber:1994uf}. Anyway, the relation is the same, that is, $n = d(\mathcal{M})$ the number of solutions at the instanton moduli space of the equations \eqref{inst1}-\eqref{inst2}. Indeed, for instanton solutions in the vicinity of $A^a_\mu$, that is, $A^a_\mu+\delta A^a_\mu$, we get from \eqref{gf3} the condition
\begin{equation} \label{inst1}
\Theta^{ab,}_{\pm,\mu\nu\beta} \delta A ^{b}_\beta = 0\;,
\end{equation}
while the Landau gauge imposes
\begin{equation}\label{inst2}
\partial_\mu \delta A^a_\mu  = 0\;.
\end{equation}
Here, $d(\mathcal{M})$ is the dimension of the moduli space $\mathcal{M}$ \footnote{For a thorough analysis of $d(\mathcal{M})$ and its relation with the first Pontryagin number of the bundle $E$ ($p_1(E)$), Euler characteristic ($\chi(M)$) and signature ($\sigma(M)$) of the manifold $M$, according to the gauge group, see \cite{Atiyah:1978wi}. For the $SU(2)$ group, for instance, $d(\mathcal{M}) = 8p_1 - \frac{3}{2}(\chi + \sigma)$.}.

As we shall discuss later, the aforementioned tree-level exactness persists when the Gribov gauge fixing ambiguity is dealt with \`{a} la Gribov-Zwanziger \cite{Gribov:1977wm,Zwanziger:1989mf,Zwanziger:1992qr}, thereby indicating that the Gribov copies are inoffensive for this type of topological theory. This then also shows that the algebraic setup of \cite{Ouvry:1988mm} remains valid, even when Gribov copies are taken into account. Before doing so, let us first briefly discuss the Gribov problem.

\section{Gribov ambiguities}

To write down the conditions for the existence of Gribov copies, i.e.,~the possibility of having multiple solutions to the gauge fixing constraints, we start with the gauge field. Let $A_{\mu}^{\prime a}$ differ from $A_{\mu}^{a}$  -- which satisfies the Landau gauge condition, by assumption -- by a pure infinitesimal gauge transformation, i.e.,~$A_{\mu}^{\prime a}=A_{\mu}^{a}+\delta A_{\mu}^{a}$; the gauge transformed field will be a copy of $A_{\mu}^{a}$ if the following is satisfied
\begin{equation}
    \partial_{\mu}A_{\mu}^{\prime a}=0\;,
\end{equation}
which amounts to
\begin{equation}
    \partial_{\mu}D_{\mu}^{ab}\omega^{b}+\partial_{\mu}\alpha_{\mu}^{a}=0\;.
    \label{eqn:gluon-top-eqn}
\end{equation}
Notice that, by virtue of the condition \eqref{gf2}, which is equivalent to saying that $\partial_\mu \alpha_\mu^a=0$, see below eq.~\eqref{eqn:F-symmetry}, the second term in the above equation actually drops out, but we will keep it for now, so that at later stage, it will become clear why the condition \eqref{gf2} is such a convenient one.

Similarly, the gauge condition \eqref{gf2} features infinitesimal copies if
\begin{equation}
    \partial_{\mu}D_{\mu}^{ab}\lambda^{b}=0\;.
    \label{eqn:top-parameter-copy-eqn}
\end{equation}
In the current context, there is the possibility for the field strength gauge condition $F_{\mu\nu}^{a}$ to have copies as well. This is a novelty introduced by the topological model, insofar as, in the usual Yang-Mills theory, $F_{\mu\nu}^{a}$ is completely defined by the first constraint on $A_{\mu}^{a}$ \eqref{gf1}, while in the topological case there is another independent gauge ambiguity involving $A^a_\mu$, which is  reflected in the behavior of the field strength that also transforms as a gauge field \eqref{eqn:F-symmetry}, as we discussed above. From the (anti-)self-dual gauge fixing \eqref{gf3}, the new condition is obtained as follows
\begin{equation}
    F_{\mu\nu}^{\prime a} \pm \tilde{ F }_{\mu\nu}^{\prime a} =  F_{\mu\nu}^{a} \pm  \tilde{ F }_{\mu\nu}^{a}\;,
\end{equation}
so that a copy is possible when
\begin{equation}
D_{[\mu}^{ab}\alpha_{\nu]}^{b} \pm \epsilon_{\mu\nu\alpha\beta} D_{\alpha}^{ab}\alpha_{\beta}^{b} = 0\;.
    \label{eqn:F-copy-eqn}
\end{equation}
In summary, the conditions for the existence of infinitesimal Gribov copies for the three local gauge parameters of the model are
\begin{align}
    \partial_{\mu}D_{\mu}^{ab}\omega^{b}+\partial_{\mu}\alpha_{\mu}^{a} &= 0\;, \label{copy1} \\
                                  \partial_{\mu}D_{\mu}^{ab}\lambda^{b} &= 0\;, \label{copy2} \\
                                     D_{[\mu}^{ab}\alpha_{\nu]}^{b} \pm
          \epsilon_{\mu\nu\alpha\beta}D_{\alpha}^{ab}\alpha_{\beta}^{b} &= 0\;. \label{cpoy3}
\end{align}
We must verify if the system of equations \eqref{copy1}-\eqref{cpoy3} allows for (normalizable) zero modes. If we set $\alpha_\mu = 0$, the third equation trivializes, while the first two reduce to
\begin{align}
    \partial_{\mu}D_{\mu}^{ab}\omega^{b} &= 0\;,\label{GribovYM1} \\
                                  \partial_{\mu}D_{\mu}^{ab}\lambda^{b} &= 0\;, \label{GribovYM2}
\end{align}
which shows that there is a sector for a particular configuration of the gauge parameters in which the usual Gribov copies are present. Indeed, these two copies equations are identical to the one which characterizes the infinitesimal Gribov problem in Yang-Mills theories in the Landau gauge \cite{Gribov:1977wm,Sobreiro:2005ec,Zwanziger:1989mf,Zwanziger:1992qr,DellAntonio:1989wae,DellAntonio:1991mms,vanBaal:1991zw}.

Analyzing the third equation separately, we can easily check that this equation also allows for zero modes. For $h(x) \in G$, we know that $h^{-1}\partial_\mu h$ belongs to the Lie algebra defined by the gauge group $G$, i.e., $h^{-1}\partial_\mu h (x)= [h^{-1} \partial_\mu h]^a(x) T^a$ where $[h^{-1} \partial_\mu h]^a $ is a scalar function for each $\mu$ (and $a$) and $T^a$ are the generators of the Lie algebra. Moreover, it is well-known that for a pure gauge configuration
\begin{equation} \label{Fh}
F_{\mu\nu}(h^{-1}\partial h) = 0\;,
\end{equation}
where $F_{\mu\nu} = F_{\mu\nu}^a T^a$. So if we set $\alpha^a_\mu = D^{ab}_\mu[h^{-1} \partial h]^b$, by using
\begin{equation}
[D_\mu, D_\nu] = F_{\mu\nu} \;,
\end{equation}
we will get in both terms of \eqref{cpoy3} the expression \eqref{Fh}, which shows in a simple way that \eqref{cpoy3} admits zero modes as well.

In the following, we discuss the relevance of these copies in view of the instanton properties of the moduli space and develop a strategy to eliminate them from the path integration.

\section{Elimination of the copies}

In order to eliminate the ambiguities related to the infinitesimal Gribov copies, we can start by eliminating the Gribov copies present in the sector $\alpha^a_\mu = 0$. For that, according to equations \eqref{GribovYM1} and \eqref{GribovYM2}, we shall implement the usual Gribov-Zwanziger restriction to the region $\Omega$ defined as \cite{Gribov:1977wm,Zwanziger:1989mf}
\begin{equation} \label{Omega}
\Omega = \{A^a_\mu; \; \partial_\mu A^a_\mu = 0, \; \mathcal{M}^{ab} > 0 \}\,,
\end{equation}
wherein
\begin{equation} \label{FPop}
\mathcal{M}^{ab}(x,y) = -\delta(x-y)\partial_\mu D^{ab}_\mu  = -\delta(x-y)(\partial^2\delta^{ab} - f^{abc}A^c_\mu\partial_\mu)\,,
\end{equation}
with $\partial_\mu D_\mu^{ab}$ depending on $y$. In a few words, one imposes that the real eigenvalues of the Hermitian operator $-\partial_\mu D^{ab}_\mu \equiv - \partial D$ are positive. At its boundary, $\partial \Omega$, the FP operator acquires its first vanishing eigenvalues. This imposition eliminates the infinitesimal copies generated by the first two equations, viz.~\eqref{copy1} and \eqref{copy2}.

Notice that we tacitly remained silent here about ``large'' Gribov copies, that are not related to FP zero modes. To deal with those, a further restriction to a subregion of $\Omega$ would be necessary, viz.~the fundamental modular region $\Lambda$ that is related to global minima of the minimizing functional $\min_{u\in \text{SU}(N)} \int d^4x A_\mu^u A_\mu^u$, whereas $\Omega$ is related to local minima. We will have nothing more to say about this \cite{vanBaal:1991zw}. As of now, it is completely unknown how to restrict in practice to $\Lambda$ that lacks a simple description as $\Omega$ in terms of \eqref{Omega}. At best, we can refer to \cite{Zwanziger:2003cf} where some partial argument---for standard Yang-Mills gauge theories---was provided that averaging over gauge configurations restricted to $\Omega$ or $\Lambda$ give coinciding expectation values for the observables of Yang-Mills theories.

In the case with $\alpha^a_\mu\neq0$ we can decompose $\alpha^a_\mu$ according to the Helmholtz decomposition \cite{Woodside:1999yj}. Since we are working in flat Euclidean space, for $\alpha^a_\mu(x)$ fields sufficiently smooth\footnote{Here the term ``sufficiently smooth" means functions that are at least $C^2$, i.e.,~twice continuously differentiable functions on the closure of the four-dimensional volume $V_4$.} that fall off as $\frac{1}{r}$ or faster at infinity, we may rely on a generalization of the Helmholtz theorem by which we can write the four-vector $\alpha^a_\mu(x)$ as
\begin{eqnarray}
\alpha^a_\mu(x) &=& - \partial_\mu \left[ \int_{V^{\prime}_4} \frac{\partial^{\prime}_\nu \alpha^a_{\nu}(x^{\prime})}{4\pi^2R^2(x,x^\prime)} d^4x^{\prime} - \oint_{\Sigma^{\prime}} \frac{\alpha^a_\nu(x^\prime) n^\prime_\nu}{4\pi^2 R^2(x,x^\prime)} d\Sigma^\prime\right] \nonumber\\
&-& \partial_\beta \left[ \int_{V^\prime_4} \frac{\partial^\prime_\beta \alpha^a_\mu(x^\prime) - \partial^\prime_\mu \alpha^a_\beta(x^\prime)}{4\pi^2 R^2(x,x^\prime)} d^4x^\prime + \oint_{\Sigma^\prime} \frac{\alpha^a_\beta(x^\prime) n^\prime_\mu - \alpha^a_\mu(x^\prime) n^\prime_\beta}{4\pi^2 R^2(x,x^\prime)} d\Sigma^\prime \right]\;,
\end{eqnarray}
with $R^2(x,x^\prime) = \vert x - x^\prime\vert^2$, and $n^\prime_\mu$ is the four-vector outward unit normal of the three-surface $\Sigma^\prime$ which encloses the four-volume $V^\prime_4$, $\Sigma^\prime$ itself being sufficiently smooth. Thus, eliminating the surface integrals for vanishing fields on the boundary according to the conditions above, we conclude that we can split $\alpha^a_\mu(x)$ into its longitudinal and transverse parts in the form
\begin{equation} \label{split}
\alpha^a_\mu = \partial_\mu \phi^a + \partial_\beta T^a_{\beta \mu}\;,
\end{equation}
where $\phi^a$ is a scalar field, and $ T^a_{\beta\mu}$ is an antisymmetric tensor given, respectively, by
\begin{equation}
\phi^a = -\int_{V^{\prime}_4} \frac{\partial^{\prime}_\nu \alpha^a_{\nu}(x^{\prime})}{4\pi^2R^2(x,x^\prime)} d^4x^{\prime}\;,
\end{equation}
and
\begin{equation}
T^a_{\beta\mu} = - \int_{V^\prime_4} \frac{\partial^\prime_\beta \alpha^a_\mu(x^\prime) - \partial^\prime_\mu \alpha^a_\beta(x^\prime)}{4\pi^2 R^2(x,x^\prime)} d^4x^\prime \;.
\end{equation}
The divergence of the second term in \eqref{split} vanishes. Therefore,
\begin{equation}
\partial_\mu \alpha_\mu^a = \partial^2\phi^a, \quad \text{where} \quad \partial_\mu \partial_\mu \equiv \partial^2.
\end{equation}
Returning to the copy equation \textcolor{red}{\eqref{copy1}}, in principle if one chooses e.g.~$\phi = -  \frac{\partial_\mu}{\partial^2}   D_\mu \omega$, then this equation \textcolor{red}{\eqref{copy1}} has a solution. This would imply, in general, that all Gribov copies that exist in Yang-Mills theories are removed, but it is logically possible to generate new ones with a non-vanishing topological shift. But now comes the fact that so far, we did not use yet the second gauge condition \eqref{gf2}. Doing so, the gauge condition for $\alpha_\mu$ (or $\psi_\mu$) demands that it must be transverse, which allows just for trivial $\phi$ (i.e., $\psi_\mu$ must be transverse). Thence, the usual Gribov restriction also eliminates the copies related to the gauge transformation of the topological parameter.

It remains to deal with eq.~\eqref{cpoy3}, the third copy equation, at a first glance, the condition $-\partial D > 0$ does not tell anything about the instantons.  We could think about an analogous procedure to eliminate the copies arising from the third equation \eqref{cpoy3}. Rewriting eq.~\eqref{cpoy3} as
\begin{equation} \label{theta}
i\Theta^{ab}_{\pm,\mu\nu\beta} \alpha^b_\beta = 0 \,,
\end{equation}
we could employ the extra Gribov-like restriction $ i\Theta^{ab}_{\pm,\mu\nu\beta}\equiv i\Theta_\pm > 0$, i.e., we would impose positive eigenvalues for the operator $i\Theta_\pm$. However, let us now motivate why this third restriction is not necessary.

Firstly, we recall that Witten noted that the partition function $Z$ of his topological theory is independent of changes of the coupling constant $g^2$ (as long as $g^2 \neq 0$). He used this liberty to compute the observables in the weak coupling limit, $g^2 \rightarrow 0$, from which he obtained the Donaldson polynomials. The evaluation of $Z$ in the weak coupling limit means that the theory is dominated by the classical minima. These minima correspond to the \mbox{(anti-)}instanton configurations $F^a_{\mu\nu} = \pm \widetilde{F}^a_{\mu\nu}$, where the $``+"$ sign corresponds to instanton, and $``-"$ to anti-instanton solutions. Once it was proven that the observables of the Witten and Baulieu-Singer theories are the same (see for instance \cite{Weis:1997kj}), we should then consider the instanton characterization not as a gauge fixing condition, but as a physical requirement in order to obtain the correct degrees of freedom  that correspond to the description of all global observables. This was also stressed in \cite{vanBaal:1989aw}: condition \eqref{gf3} does not completely fix the gauge, on purpose, to be left with the finite set of degrees of freedom describing the instantons, the latter being exactly the kernel of \eqref{gf3}. In fact, the (bosonic) ``zero modes of the 3rd kind'' will be exactly cancelled in computations against fermionic zero modes, related to the $\bar\chi$-equation of motion, see again \cite{vanBaal:1989aw}. Precisely, the Atiyah-Singer index theorem \cite{Atiyah:1963zz} counts the number of solutions of \eqref{inst1} and \eqref{inst2}, which gives the correct dimension of the instanton moduli space, in complete harmony with instanton conformal properties \cite{Witten:1976ck,Jackiw:1976fs}. In this sense, the structure of \eqref{inst1} and \eqref{inst2}, and therefore of \eqref{theta}, are protected by the Atiyah-Singer theorem and its direct correspondence with the conformal properties of instanton configurations, indicating that no extra physical restrictions on the eigenvalues of $i\Theta_\pm$ need to be introduced.

However, one might question whether the restriction of the gauge fields to the Gribov region does not hamper the fact that we wish to ``preserve'' the instantons, as just motivated.  In the case of the simplest $SU(2)$ instanton, we can provide an affirmative answer to this, inspired by the observations of \cite{Maas:2005qt}. Indeed, in this case the instanton field with winding number $1$ is given by the expression (see e.g.~\cite{Bohm:2001yx})
\begin{equation}
{A^\text{(i)}}^a_\mu = \frac{1}{g} \frac{2}{r^2+ \lambda^2} r_\nu \zeta^a_{\nu\mu}\;,
\end{equation}
where $\lambda$ denotes the $``$size" of the instanton, while the real constant antisymmetric matrices $\zeta^a$ are the 't Hooft tensors that obey the algebra
\begin{eqnarray}
\left[\zeta^a, \zeta^b\right] &=& 2f^{abc}\zeta^c\;, \nonumber \\
\{\zeta^a, \zeta^b\} &=& -\delta^{ab}\;.
\end{eqnarray}
As we can see,
\begin{equation}
\partial_\mu {A^{a}_\mu}^{(i)}=0\;,
\end{equation}
which means that the (regular) instanton field is transverse and in the Landau gauge. From the latter transversality of the instanton field, the eigenvalue equation for the FP operator \eqref{FPop},
\begin{equation}
\mathcal{M}^{ab}(A^{(i)}) \phi^a = -\omega^2 \phi^a,
\end{equation}
takes the form
\begin{equation} \label{eigen}
\partial^2 \phi^a + f^{abc}   \frac{2}{r^2+ \lambda^2} r_\mu \zeta^a_{\nu\mu}  \partial_\nu \phi^c = -\omega^2 \phi^a.
\end{equation}
We immediately notice that this instanton has three trivial constant zero-modes. The other zero modes (thus giving $\omega = 0$) of eq.~\eqref{eigen} were explicitly constructed in \cite{Maas:2005qt}. This means that the instanton belongs to the Gribov horizon $\partial\Omega$.

There is no strict proof that all instantons (with higher winding number) belong to the first Gribov region, but to the best of our knowledge, in the cases investigated in literature, topological Yang-Mills solutions (instanton, monopole, vortex) always belong to it---see again \cite{Maas:2005qt}, or \cite{Bruckmann:2000xd} for an example in the Maximal Abelian gauge. Let us also refer to \cite{deForcrand:2000yr}, where it was discussed that for instantons a whole family of Gribov copies does exist.

The consequence of such rich zero-mode spectrum to our problem is immediate. If we consider the Gribov restriction, $-\partial D>0$, for a generic gauge field in order to eliminate the Gribov copies in the first two copies equations, \eqref{copy1} and \eqref{copy2}, the instantons belongs to the boundary of the first Gribov region, $\partial\Omega$ (where $-\partial D$ becomes zero) and are as such not eliminated from the game. One notes this property by the fact that the instantons are transverse, and the spectrum of the FP operator evaluated for an instanton displays zero modes. From the point of view of gauge copies under $-\partial D\geq0$, the gauge fields obeying the (anti-)self dual condition $F = \pm \widetilde{F}$ are well-defined. The solutions to $F = \pm \widetilde{F}$ are elements of $\partial \Omega$.

The Gribov problem can also be directly understood from the partition function related to the action \eqref{Stotal}. We use the expression \eqref{Zfp5}, inserted into \eqref{Zgf}. We can then also integrate out the $c$ and $\bar c$ to get the Faddeev-Popov determinant $\det(-\partial D)$.  Doing so, we arrive at the following partition function
\begin{equation}\label{part}
Z=\int\;\mathcal{D}A\mathcal{D}\psi\;\delta(\partial A)\det(-\partial D)\delta(F\pm \tilde F)\delta(\partial\psi)\det(\Theta_\pm)\exp\left\{-S_0[A]\right\}\,.
\end{equation}
Consider now (normalizable) zero modes $\xi_1$ and $\xi_2$ of the Faddeev-Popov operator,
\begin{equation}
\label{zeromodes}
 \partial_\mu D_{\mu}^{ab} \xi_1^b=\partial_\mu D_{\mu}^{ab} \xi_2^b=0
\end{equation},
and consider the following field variations (see of course \eqref{eqn:gluon-symmetry}-\eqref{eqn:top-parameter-symmetry})
\begin{eqnarray}\label{nieuwetransfo}
 \hat\delta A_\mu^a &=& D_\mu^{ab}\xi_1^b+\hat\alpha_\mu^a\,,\nonumber\\
 \hat\delta \hat\alpha_\mu^a &=&  D_\mu^{ab}\xi_2^b\,,
\end{eqnarray}
where $\hat\alpha_\mu^a$ is subject to $\partial_\mu\hat\alpha_\mu^a=0$. First of all, setting $A'=A+\hat\delta A$, we have $\partial A'=0$, so the $\delta$-function in \eqref{part} is $\hat\delta$-invariant. The associated Jacobian is trivial, i.e.~the integration measure does not change. Indeed, the shift over $\hat\alpha$ is irrelevant for the Jacobian, while for the first piece, we may use the classical argument as why the integration measure over a gauge field is gauge invariant, this mainly by virtue of the anti-symmetry of the $f^{abc}$, see e.g.~\cite{weinberg1996quantum}.  The classical action $S_0[A]$ is also invariant under \eqref{nieuwetransfo} as this is a special case of \eqref{eqn:gluon-symmetry}-\eqref{eqn:top-parameter-symmetry}. Moreover, if $F\pm \tilde F=0$, then also $F'\pm \tilde F'=0$ as $\hat\delta F_{\mu\nu}^{a}=-gf^{abc}\xi^{b}F_{\mu\nu}^{c}$. Nextly, as it is known, the Faddeev-Popov determinant is (perturbatively) gauge invariant under $A_\mu\to A_\mu+D_\mu\omega$, see e.g.~\cite{Pokorski:1987ed}. More precisely, one has $\Delta[A]\equiv[\det(-\partial D)]^{-1}=\int \mathcal{D}g~ \delta (\partial A^g)$
where $g$ is a generic SU($N$) transformation. It is then easy to show that $\Delta[A^{g'}]=\Delta[A]$. Indeed, as $\partial\hat\alpha=0$, the previous argument is unaffected by the extra shift over $\hat\alpha$ defining $\hat\delta A$. Overall, the $\det(-\partial D)$ will thus be untouched by the transformation generated by \eqref{nieuwetransfo}. To establish full invariance of $Z$, we just need to show that also $\det(\Theta_\pm)$ has a trivial variation. That this is the case can be realized from writing\footnote{This is completely analogous to the Faddeev-Popov ``unity trick'' and the proof that the corresponding determinant is perturbatively gauge invariant.}
\begin{eqnarray}\label{nieuwargument}
1 = \det(\Theta_\pm)\int \mathcal{D}U~\delta(F^U\pm \tilde F^U)\,,
\end{eqnarray}
where $U$ is a generic transformation generated by the infinitesimal topological shift $\delta_{shift} A= \zeta$ for arbitrary $\zeta$.  From this, the (perturbative) invariance of $\det(\Theta_\pm)$ under this topological shift follows from the invariance of the measure. A fortiori, $\det(\Theta_\pm)$ will then also be $\hat\delta$-invariant which is just a special case upon setting $\zeta=D\xi_1+\hat\alpha$.

Eventually, we thus see that even after gauge fixing, the zero modes of the Faddeev-Popov operator, \eqref{zeromodes}, induce still an overcounting of the relevant degrees of freedom, encoded in the $\hat\delta$-invariance of the gauge fixed partition function \eqref{part}. The foregoing reasoning also shows why it is sufficient to restrict to the ``standard'' Gribov horizon known from usual Yang-Mills theories, given the form of the residual $\hat\delta$-invariance.

Summing up, the only requirement to eliminate all (infinitesimal) gauge ambiguities is then the introduction of the Gribov horizon as it commonly done for usual Yang-Mills theories\footnote{Although all points discussed here indicate a similar behavior for a generic $SU(N)$ instanton field with an arbitrary winding number, a possible analytical treatment of such instantons will not be considered in this paper.}. Then it remains to prove in the following section that also this restriction to the standard Gribov horizon eventually becomes trivial at the dynamical level.

\section{The Gribov gap equation and its triviality}

We have mentioned that the tree-level exactness of the topological theory in the (A)SDL gauges \cite{Junqueira:2018xgl} suggests that the Gribov copies present in our model should be inoffensive. Due to the absence of radiative corrections, the tree-level propagator of the FP ghost field in momentum space obtained from the total action \eqref{Stotal},
\begin{equation}
\langle \bar{c}^a c^b \rangle_0\left(p\right) = \delta^{ab} \frac{1}{p^2}\;,
\end{equation}
will be valid to all orders in perturbation theory. From the expression above, one sees that the FP operator will be positive definite at the quantum level, consistent with the inverse of the FP propagator being positive, i.e., we are inside the first Gribov region, in such a way that the Gribov restriction to the path integral seems to be redundant. The origin of such behavior is the impossibility of closing loops in Feynman diagrams, as due to the vertex structures, at least one gauge field propagator is required to close loops, but $\langle A_\mu^a(x) A_\nu^b(y) \rangle=0$ to all orders for this gauge choice \cite{Junqueira:2018xgl,Junqueira:2018zxr}. We point out that the same argument holds for the analysis of the third Gribov equation \eqref{cpoy3} and the propagator $\langle\bar{\chi}^a_{\mu\nu}\psi_\alpha^b\rangle_0(p)$.

Originally, the no-pole condition was achieved by treating the gauge field as an external source. Its quantum properties must be computed when the gauge field is integrated over. If we admit the Gribov copies to play a role in this case, we should consider that the introduction of the term that implements the restriction to the Gribov region might allow for radiative corrections, e.g.~from a non-vanishing gauge propagator arising from the extra Gribov term (a metric dependent term) in the action. This might perturb the original cohomology arguments and, consequently, compromise the global properties of the topological theory at certain energy scale,  this through the elimination of Gribov ambiguities. Taking into account the reasons discussed above, such behavior is highly unexpected. We will now show this in detail, first at one loop, afterwards we will generalize to all orders.

\subsection{No-pole condition at one-loop}

As discussed, all infinitesimal Gribov copies in the topological theory in (A)SDL gauges for the $SU(2)$ instanton are eliminated through the implementation of the restriction to the well-known Gribov region denoted by $\Omega$ \eqref{Omega}, commonly performed in usual Yang-Mills theories in Landau gauge. Following the Gribov approach, this restriction is achieved via the introduction of a form factor $V(\Omega)$ in the generating function $Z[J]$, in such a way that the integration domain is limited by $\Omega$. The original generating functional
\begin{equation}
Z_o[J] = \mathcal{N} \int \mathcal{D}\Phi \, e^{-S - \int d^4x J\Phi}\, ,
\end{equation}
is restricted to
\begin{equation} \label{Z}
Z[J] =\mathcal{N} \int_{\Omega} \mathcal{D}\Phi \, e^{-S -\int d^4x J\Phi} =\mathcal{N} \int \mathcal{D}\Phi V\left(\Omega\right)e^{-S-\int d^4x J\Phi} \;,
\end{equation}
where $\mathcal{N}= Z[0]^{-1}$ is the normalization factor, $\mathcal{D}\Phi$ denotes the integration measure for all fields, i.e., $\mathcal{D}\Phi = \mathcal{D}A \mathcal{D}\psi \mathcal{D}c \mathcal{D} \phi \mathcal{D} \bar{c} \mathcal{D} b \mathcal{D} \bar{\phi} \mathcal{D} \bar{\eta} \mathcal{D} \bar{\chi} \mathcal{D} B $, while $J\Phi = J_i \Phi_i$ denotes the coupling of each field $\Phi_i$ with its respective external source $J_i$.

In the Yang-Mills theory, the form factor $V(\Omega)$ is obtained from the no-pole condition for the FP propagator, since the imposition $\mathcal{M}^{ab} > 0$ is equivalent to forbidding the existence of poles in the FP propagator \cite{Gribov:1977wm,Zwanziger:1989mfb}. In  the topological case, see action \eqref{S_gf}, the operator $\mathcal{M}^{ab} = -\partial_\mu D_\mu^{ab}$ appears twice: in the FP ghost-anti-ghost quadratic term (treating $A^a_\mu$ as an external source), $\bar{c} \partial D c$, as usual, but also in the bosonic ghost-anti-ghost term, $\bar{\phi} \partial D \phi$. By applying the Gribov semi-classical method we shall see that, at one-loop order, the no-pole condition in the topological theory takes the same form as for the standard Yang-Mills case.

For this purpose, we have only to analyze the vertices present in the total action \eqref{Stotal}, and apply the Feynman rules for the diagrams up to the order $g^2$, once we are considering the one-loop order. We should then verify which diagrams could be constructed with an incoming $\bar{c}$-leg ($\bar{\phi}$-leg), and an outgoing $c$-leg ($\phi$-leg), whereby the gauge fields work as external sources. Let us start with the FP ghost propagator.

(i) \textit{FP ghost propagator}. Using the following notation for the ghost propagator at one-loop with $A$ as an external source,
\begin{equation} \label{ghostprop}
\langle \bar{c}^a(k) c^b(p) \rangle = \delta(p+k)\mathcal{G}^{ab}(k^2, A) = \delta(p+k) \delta^{ab} \frac{1}{k^2}\left[1+\sigma(k,A)\right],
\end{equation}
our aim is to calculate $\sigma(k,A)$, which represents the loop correction to the tree-level part $1/{k^2}$. Firstly, we must note that the FP anti-ghost, $\bar{c}$, only propagates to $c$ and $\psi$  through the propagators $\langle \bar{c} c \rangle_0$ and $\langle \bar{c} \psi \rangle_0$ at the tree-level, respectively. Therefore, if we start with an incoming $\bar{c}$, we can propagate it to the vertices (a) $\bar{\phi} c \psi$, (b) $\bar{\chi} \partial A \psi$, (c) $\bar{\chi} c A$, (d) $\bar{\chi} c AA$, or (e) $\bar{c} A c$. The first one does not produce external $A$-legs. If we propagate $\bar{c}$ to the vertex (b) through $\langle \bar{c}\psi \rangle_0$, we will get an external $A$-leg, and an internal $\bar{\chi}$-leg. Since $\bar{\chi}$ only propagates to $\psi$ through $\langle \bar{\chi} \psi \rangle_0$, we could only connect at one-loop order the vertex (b) to another vertex $\bar{\chi} \partial A \psi$, producing one more time an external $A$-leg, and an internal $\bar{\chi}$, in such a way that we cannot generate an outgoing $c$. For the vertices (c) and (d), we fall back to the same situation: we generate external $A$-legs, but always accompanied by the internal $\bar{\chi}$-leg that never propagates to $c$ in the end. We conclude that the only possibility to get an outgoing $c$ from $\bar{c}$ with only external $A$-legs is to construct the diagram by using the vertex (e)\footnote{We remark that the whole argument can be made easier by a redefinition $\bar{\eta}\mapsto\bar{\eta}+\bar{c}$ in the action \eqref{Stotal} in order to eliminate the $\bar{\eta}\psi$ mixing term.}. Namely, for
\begin{equation} \label{G}
\mathcal{G}(k^2,A) = \frac{1}{N^2-1} \delta^{ab}G(k^2,A)^{ab}\;,
\end{equation}
we construct the diagrams
\begin{small}
\begin{equation}
\mathcal{G}^{ab}(k^2, A) =
\begin{tikzpicture}
  \begin{feynhand}
    \vertex (a1) {\(\overline{c}^a\)};
    \vertex[right=1.7cm of a1] (a2) {\(c^b\)};
    \graph {
      {[edges=fermion]
        (a1) -- (a2),
        },
            };
    \end{feynhand}
\end{tikzpicture}
+
\begin{tikzpicture}
  \begin{feynhand}
    \vertex (a1) {\(\overline{c}^a\)};
    \vertex[right=1.65cm of a1] (a2);
    \vertex[right=1.65cm of a2] (a3) {\(c^b\)};
    \vertex[above=1.65cm of a2] (b1){\(k,\mu\)};
  \graph {
  {[edges=fermion]
    (a1) -- [fermion, edge label=\(k\)]  (a2),
        (a2) -- [fermion, edge label=\(p\)](a3),
        },
   (a2) -- [gluon, edge label=\(k-p\)] (b1)
   };
   \end{feynhand}
\end{tikzpicture}
+
\begin{tikzpicture}
  \begin{feynhand}
    \vertex (a1) {\(\overline{c}^a\)};
    \vertex[right=1.65cm of a1] (a2);
    \vertex[right=2.4cm of a2] (a3);
    \vertex[right=1.65cm of a3] (a4) {\(c^b\)};
    \vertex[above=1.65cm of a2] (b1){\(k,\mu\)};
    \vertex[above=1.65cm of a3] (b2){\(l,\nu\)};
    \graph {
      {[edges=fermion]
        (a1) -- [fermion, edge label=\(k\)]  (a2),
        (a3) -- [fermion, edge label=\(p\)](a4),
      },
      (a2) -- [fermion, edge label=\(k+p^\prime\)] (a3),
      (a2) -- [gluon, edge label=\(-p^\prime\)] (b1),
      (a3) -- [gluon, edge label=\(p^\prime + k -q\)] (b2)
    };
    \end{feynhand}
\end{tikzpicture}
.\label{fig}
\end{equation}
\end{small}
The possible diagrams are reduced to the same diagrams of the standard Yang-Mills theory. We conclude that the no-pole condition for the FP ghost propagator in this topological model gives the same result of the one found for Yang-Mills theory. The Feynman rule for the vertex $\bar{c}^a\partial_\mu A^k_{\mu} c^b$ is given by $ik_\mu f^{akb}$, where the incoming momentum $k_\mu$ stems from $\bar{c}$. Hereafter, just to remember, these diagrams represent, in $d$ dimensions, the three integrals below
\begin{eqnarray}
I_1 &=& \delta^{ab} (2\pi)^d \delta(k-q) \frac{1}{k^2} \;,\label{I1}\\
I_2 &=& g \frac{1}{k^2}\frac{1}{p^2} f^{akb} ip_\mu A^k_\mu(k-p)\;,\label{I2}\\
I_3 &=& g^2 \int \frac{d^d p^\prime }{(2\pi)^d} \frac{1}{k^2} \frac{1}{\left(p^\prime +k\right)^2} \frac{1}{p^2} f^{akc} i \left(p^\prime  + k_\mu \right) A^k_\mu(-p^\prime)f^{c\ell b} iq_\nu A^\ell_\nu(p^\prime+k-q)\;. \label{I3}
\end{eqnarray}
As it is known \cite{Gribov:1977wm,Sobreiro:2005ec}, we must disregard $I_2$. Due to the vertex and propagator structure of the total action \eqref{Stotal}, there is no way to close loops from the second diagram after integrating over the gauge field. Replacing \eqref{I1} and \eqref{I3} into $\mathcal{G}(k^2,A)$ \eqref{G} yields
\begin{equation}
\mathcal{G}(k^2,A) = \frac{1}{k^2} + \frac{Ng^2}{k^4\left(N^2-1\right)V} \int \frac{d^dq}{(2\pi)^d} A^a_\mu(-q)A^a_\nu \frac{(k-q)_\mu q_\nu}{(k-q)^2}\;,
\end{equation}
therefore, from \eqref{ghostprop},
\begin{equation} \label{sigmakA}
\sigma(k,A) =  \frac{Ng^2}{k^2\left(N^2-1\right)V}  \int \frac{d^dq}{(2\pi)^d} A^a_\mu(-q)A^a_\nu \frac{(k-q)_\mu q_\nu}{(k-q)^2} \;,
\end{equation}
wherein $V$ is the infinite volume factor, and $N=2$ as we are working with $SU(2)$. For small $\sigma(k^2,A)$, the Born approximation may be employed,
\begin{equation}
\mathcal{G}(k^2,A) \sim \frac{1}{k^2} \frac{1}{1- \sigma(k,A)}\;,
\end{equation}
the no-pole condition that corresponds to the restriction of the domain to the Gribov region reads
\begin{equation}
\sigma(k,A) < 1\;.
\end{equation}
As $\sigma(k,A)$ decreases for increasing $k^2$ (see \cite{Kroff:2018ncl}), the condition above is equivalent to imposing
\begin{equation} \label{no-pole}
\sigma(0,A) < 1\;,
\end{equation}
where, taking the limit $k^2 \rightarrow 0$ in \eqref{sigmakA},
\begin{equation} \label{sigma}
\sigma(0,A) = \frac{g^2 N}{4V (N^2-1)} \int \frac{d^4q}{(2\pi)^4} \frac{A^a_\lambda(q) A^a_\lambda(-q)}{q^2}\;,
\end{equation}
which defines the form factor $V(\Omega)$ as the theta function\footnote{$\Theta(x) = 1$ if $x>0$, $\Theta(x) = 0$ if $x<0$.}
\begin{equation}
V(\Omega) = \Theta\left(1-\sigma(0,A)\right)\;,
\end{equation}
or, using the Heaviside expression,
\begin{equation} \label{V}
V(\Omega) = \int^{+i\infty+\epsilon}_{-i\infty+\epsilon} \frac{d\xi^2}{2\pi i \xi^2} e^{\xi^2\left(1-\sigma(0,A)\right)}\;.
\end{equation}
We should then introduce this factor into the path integral in order to implement the elimination of the gauge copies. We must do the same procedure to eliminate possible copies in the bosonic ghost propagator, but as we will see now, the no-pole condition \eqref{no-pole} for the bosonic ghost is valid for both, the FP and bosonic ghosts.

(ii) \textit{Bosonic ghost propagator}. The proof of the last statement is immediate. The bosonic anti-ghost field $\bar{\phi}$ only propagates to $\phi$ through $\langle \bar{\phi} \phi \rangle_0$, thus an incoming $\bar{\phi}$, we can only connect to the vertex $\bar{\phi} A \phi$. Aftermath, the construction of the Feynman diagrams up to $g^2$ order with $A$ fields as external sources takes the same form of the FP case, see \eqref{fig}, only replacing $\bar{c}$ by $\bar{\phi}$, and $c$ by $\phi$. The Feynman rules are exactly the same, consequently the no-pole condition for the bosonic ghost generates the same expression for $\sigma(k,A)$, and the condition \eqref{no-pole} is valid for the FP and bosonic ghost sectors.

In a few words, although the complex structure of the total action \eqref{Stotal}, in which there are two ghosts sectors to implement the no-pole condition, for the FP ghost sector and the bosonic one, the elimination of all Gribov copies in the topological Yang-Mills in the (A)SDL gauges for $SU(2)$ instantons is achieved by introducing in the path integral a form factor $V(\Omega)$ \eqref{V} which is identical to the one obtained in the usual Yang-Mills theory in the Landau gauge.

\subsection{Gap equation at one-loop}
From \eqref{Stotal}, \eqref{Z} and \eqref{V}, the generating functional for the first Gribov region takes the form
\begin{equation}\label{part}
Z = N \int \mathcal{D} A^a_\mu \mathcal{D}{\Phi^\prime} \int \frac{d\xi^2}{2\pi\xi^2 i} \exp\{\xi^2 - S - \xi^2\sigma(0,A)\}.
\end{equation}
in which $\Phi^\prime$ denotes all fields except the gauge field. The effective potential, $\Gamma$, is defined as usual by
\begin{equation}
e^{-\Gamma} = e^{-V \varepsilon} = Z\;,
\end{equation}
where $\varepsilon$ represents the vacuum energy.

In order to calculate $\Gamma$ at one-loop order, $\Gamma^{(1)} = V \varepsilon^{(1)}$, we must select only the quadratic part of the total action $S$ (here $\sigma(0,A)$ is already quadratic as it was only calculated up to one-loop order), namely,
\begin{equation}
e^{-V\varepsilon^{(1)}} = Z_{quad}\;,
\end{equation}
whereby, using \eqref{sigma},
\begin{equation}
e^{-\Gamma^{(1)}} = \int \mathcal{D} A^a_\mu \mathcal{D}{\Phi^\prime} e^{-S_{\text{quad}}[\Phi]}.
\end{equation}
After integrating out the auxiliary fields $b^a$, $B^a_{\mu\nu}$, and all other fields except $A_\mu^a$, we get the quadratic action for the gauge field
\begin{equation}
S_{\text{quad}}[A] = \int d^4p A^a_{\mu}(p) \left[ \frac{4}{\beta} p^2 \delta_{\mu\nu} - \left(\frac{4}{\beta} - \frac{1}{\alpha}\right)p_\mu p_\nu \right]A^a_\nu(-p)+\text{rest}\;.
\end{equation}
Taking into account all quadratic terms,
\begin{equation}\label{qq}
Z_{quad} = N \int \mathcal{D} A^a_\mu \int \frac{d\xi^2}{2\pi\xi i} \exp\left\{\xi^2 -\ln{\xi} - \frac{1}{2}\int \frac{d^4k}{(2\pi)^4} A^a_\mu(k) Q_{\mu\nu}(k, \xi)\delta^{ab} A^b_\nu(-k) + \text{rest}\right\}\;,
\end{equation}
wherein
\begin{equation}
Q_{\mu\nu}(k, \xi) = \left[\frac{4}{\beta}k^2 + \frac{\xi^2 g^2 N}{2V(N^2-1)k^2}\right] \delta_{\mu\nu} + \left(\frac{1}{\alpha} - \frac{4}{\beta}\right)k_\mu k_\nu\;.
\end{equation}
Therefore,
\begin{equation} \label{Zquad}
Z_{quad} = N \int \frac{d\xi}{2\pi i} e^{[f(\xi) + \text{rest}^\prime]}\;,
\end{equation}
where,
\begin{equation} \label{f}
f(\xi) = \xi^2 - \frac{1}{V}\ln\xi + \ln[(\det Q_{\mu\nu}\delta^{ab})^{-\frac{1}{2}}] = \xi^2 - \frac{1}{V}\ln\xi - \frac{1}{2}\ln\det [Q_{\mu\nu}\delta^{ab}]\;.
\end{equation}
We also changed the variable $\xi^2 \rightarrow \xi^2V$ to pull out explicitly the volume factor here, to make clear that the action is an extensive quantity ($\sim V$).

Calculating the determinant, we find
\begin{equation}\label{lndetQ}
\ln\det[{Q_{\mu\nu}}\delta^{ab}] = (N^2-1)(d-1) \sum_k \ln \left( \frac{\beta A + 4k^4}{\beta k^2}\right) + (N^2-1) \sum_k \ln \left( \frac{k^2}{\alpha} + \frac{A}{k^2}\right)\;,
\end{equation}
where
\begin{equation} \label{A}
A = \frac{\xi^2 g^2 N}{2(N^2-1)},
\end{equation}
and $k$ refers to momenta in Fourier space. Working out the last term of \eqref{lndetQ}, we get
\begin{eqnarray}
\sum_k \ln \left( \frac{k^2}{\alpha} + \frac{A}{k^2}\right) = \sum_k \ln \left(\frac{k^4}{\alpha} + A\right) - \sum_k \ln k^2\,,
\end{eqnarray}
Taking the thermodynamic limit and employing dimensional regularization, the last term vanishes. Therefore
\begin{eqnarray}
\sum_k \ln \left( \frac{k^2}{\alpha} + \frac{A}{k^2}\right) = V \int \frac{d^d k}{(2\pi)^d} \ln \left(\frac{k^2}{\sqrt{\alpha}} + i \sqrt{A}\right) + V\int \frac{d^d k}{(2\pi)^d} \ln \left(\frac{k^2}{\sqrt{\alpha}} - i \sqrt{A}\right) \sim \alpha^{\frac{d}{4}}.
\end{eqnarray}
In the limit $\alpha \rightarrow 0$, this term also vanishes. In the end,
\begin{equation}
\ln\det[{Q_{\mu\nu}}\delta^{ab}] = (N^2-1)(d-1) \int \frac{d^d k}{(2\pi)^d} \ln \left( \frac{\xi^2 g^2 N}{2(N^2-1)k^2} + \frac{4k^2}{\beta}\right),
\end{equation}
which could be rewritten as
\begin{equation}
\ln\det[{Q_{\mu\nu}}\delta^{ab}] = (N^2-1)(d-1) \left[\int \frac{d^d k}{(2\pi)^d} \ln\left( \beta\xi^2 g^2 N + 4k^2 \right) -  \int \frac{d^d k}{(2\pi)^d} \ln \left( 2\beta(N^2-1)k^2 \right)\right].
\end{equation}
In dimensional regularization, not only the last term is zero, but also the first one, as we should still take the limit $\beta \rightarrow 0$. We conclude that
\begin{equation}
f(\xi) = \xi^2\;,
\end{equation}
as we work in the thermodynamic limit, $V\to\infty$. The gap equation, viz.~the equation for the critical point for a saddle point evaluation, thus gives the trivial solution
\begin{equation}\label{sadd}
\xi = 0\;,
\end{equation}
to $f'(\xi)=0$. So, up to leading order, the no-pole condition does not change the partition function at all, see  \eqref{part} in conjunction with \eqref{sadd}.

\subsection{Extension to all orders}
Let us now extend the result \eqref{sadd} and prove that is valid to all orders in perturbation theory. Therefore, we will rely on the local version of the horizon function. Following the steps of \cite{Zwanziger:1989mf,Zwanziger:1992qr}, the restriction to the region $\Omega$ to all orders is given by considering the following partition function,
\begin{equation}
Z= \int \mathcal{D} {\Phi}
\mathrm{e}^{-S
+\gamma^4H(A)-4V\gamma^4(N^2-1)}\,,
\label{z2}
\end{equation}
where $S$ is defined in \eqref{Stotal} and $H(A)$ is the Zwanziger horizon function,
 \begin{equation}\label{zh}
H(A)=g^2\int d^4x d^4y~f^{abc}A^{b}_{\mu}(x)\left[{\cal M}^{-1}\right]^{ad}(x,y)f^{dec}A^{e}_{\mu}(y).
\end{equation}
Notice that $H(A)$ reduces to $\frac{g^2N}{V} \int d^dx A^a_\mu(x) \frac{1}{\partial^2} A^a_\mu(x)$ at lowest order, in fact recovering $\sigma(0,A)$ of the no-pole condition at one-loop \eqref{sigma}\footnote{About the exact equivalence between the no-pole ghost condition and the Zwanziger function, see for instance \cite{Gomez:2009tj,Capri:2012wx}.}. In the all-order Gribov-Zwanziger formalism, the $\Theta$-function is also replaced by a $\delta$-function in the thermodynamic limit \cite{Zwanziger:1989mf,Zwanziger:1992qr}, $V \rightarrow \infty$, as we have made explicit before.

The non-local horizon function $H(A)$ can be equivalently written in a local form through a pair of bosonic auxiliary fields  $(\bar{\varphi},\varphi)^{ab}_\mu$ and a pair of anticommuting fields $(\bar{\omega},\omega)^{ab}_\mu$ \cite{Zwanziger:1992qr}. In the current case, it means replacing the exponent of \eqref{z2} by the local action
\begin{equation}
S_{loc} = S-\int d^4x\left(\bar{\varphi}^{ac}_{\mu}{\cal M}^{ab}(A) {\varphi}^{bc}_{\mu}-\bar{\omega}^{ac}_\mu{\cal M}^{ab}(A) \omega^{bc}_\mu
+\gamma^{2} ~gf^{abc}A^{a}_{\mu}(\varphi+\bar{\varphi})^{bc}_\mu \right) \;.
\label{intro10}
\end{equation}
In the local formulation, the gap equation reads \cite{Zwanziger:1992qr}
\begin{equation} \label{gap}
\frac{\partial \varepsilon}{\partial \gamma^2} = 0\;.
\end{equation}
This relation connects the semi-classical method characterized by the no-pole ghost condition with the Zwanziger horizon function. Indeed, for the reader's belief, let us analyze the leading order limit.

At one-loop order, the geometric interpretation of thermodynamic limit is very simple: the Gribov no-pole condition \eqref{no-pole}, replacing $\frac{A^a_\mu(k)}{\sqrt{k^2}}$ by ${x^a_\mu}_{\overrightarrow{k}}\equiv x_{\overrightarrow{k}}$, could be written as
\begin{eqnarray} \label{hyper}
\frac{1}{V}\sum_{\overrightarrow{k}} x_{\overrightarrow{k}} x_{-\overrightarrow{k}} < r^2 \;,
\end{eqnarray}
where $r^2 =\frac{4(N^2-1)}{g^2}$. The expression above can be interpreted as an hypersphere in an infinite dimensional space. As it is well-known for hyperspheres, as the dimension grows, the volume of a hypersphere is getting more and more concentrated on the boundary, i.e.,~on the hypersurface defined, in our case, by the ellipsoid
\begin{eqnarray} \label{ellipsoid}
\frac{1}{V}\sum_{\overrightarrow{k}} x_{\overrightarrow{k}} x_{-\overrightarrow{k}} = r^2 \;,
\end{eqnarray}
which means that the $\Theta$-function that represents \eqref{hyper} could effectively be replaced by a $\delta$-function in the thermodynamic limit. The collapse of the $\Theta$-function into the $\delta$-function is then expressed by
\begin{equation}
\int^{+i\infty+\epsilon}_{-i\infty+\epsilon} \frac{d\xi^2}{2\pi i \xi^2} e^{\xi^2\left(1-\sigma(0,A)\right)} \rightarrow \int^{+i\infty+\epsilon}_{-i\infty+\epsilon} \frac{d\xi^2}{2\pi i} e^{\xi^2\left(1-\sigma(0,A)\right)}= \int^{+\infty}_{-\infty} \frac{d\xi^2}{2\pi} e^{i\xi^2\left(1-\sigma(0,A)\right)}\;,
\end{equation}
after a Wick rotation. In practice we just canceled the $\xi^2$ in the denominator, responsible for the second term in \eqref{f}. The behavior of $\xi$, in turn, is only determined by the gap equation \eqref{ellipsoid}. The vacuum energy can be computed from the $\ln \det$ originating from the action \eqref{intro10}, leading to exactly the same result as in the previous subsection, upon identifying $\xi^2$ and $\gamma^4$.

We conclude, without inconsistency between the both methods, that the Gribov copies (still at one-loop so far) are inoffensive to the $SU(2)$ topological Yang-Mills theory in the (A)SDL gauges, since the gap equation forbids the introduction of a Gribov massive parameter in the thermodynamic limit,
\begin{equation}
\xi \sigma(0,A) \sim \gamma^4 \int d^4k A \frac{1}{k^2} A \rightarrow 0 \;.
\end{equation}

Finally, let us look to what happens beyond the $\ln \det$-level. Then, the vertices of the theory will start to play role. Based on the vertex structure of $S_{loc}$, it is easy to see that any vacuum diagram beyond one-loop will contain at least one $\braket{AA}$-propagator. However, by inverting the quadratic form in \eqref{qq}, this propagator is given by\footnote{For the reader's convenience, we have listed all propagators in Appendix A. Moreover, since the proof discussed in this section is based on the absence of radiative corrections \cite{Junqueira:2018xgl}, this appendix is dedicated to the proof that such properties remain valid for the inclusion of the horizon function.}
\begin{equation}
\;D^{ab}_{\mu\nu} = \delta^{ab} \left[ \frac{\beta}{4}\frac{p^2}{\left(p^4+\beta N g^2\gamma^4/2\right)}\left(\delta_{\mu\nu}-\frac{p_\mu p_\nu}{p^2}\right) + \alpha\frac{p^2}{\left(p^4+2\alpha Ng^2\gamma^4\right)} \frac{p_\mu p_\nu}{p^2}\right]\;,
\end{equation}
i.e.,
\begin{equation} \label{AA=0}
\langle AA \rangle = 0
\end{equation}
if we take $\alpha$, $\beta \rightarrow 0$, irrespective of $\gamma^2$. We immediately get that all higher order terms to the vacuum energy vanish, just as the $\ln \det$. As such, by employing the gap equation \eqref{gap} which is valid to all orders, we can infer that the massive Gribov parameter vanishes to all orders in the thermodynamic limit. In this way, the global (topological and cohomological) properties of the original action are not violated and we come to the main result of this paper: quantization of the topological theory remains valid as it is, the resolution of the infinitesimal (``small'') Gribov copy problem is trivial as the intrinsic topological features of the theory self-consistently forbid the introduction of the Gribov mass, the crux of the Gribov-Zwanziger restriction \cite{Gribov:1977wm,Zwanziger:1989mf,Zwanziger:1992qr} when it comes to changing the structure of the theory.

One might wonder if it actually makes sense to have computed the above effective action by expanding around the trivial $A=0$ sector, thinking about the importance of the instanton configurations for topological field theories.

Exactly the topological nature of the theory saves the day here. Let us first remark that it is possible to write down a BRST invariant version of the Gribov restriction, that is, if $\gamma$ were to be nonzero, whilst preserving equivalence with the above formalism\footnote{In the sense that all correlation functions will be identical.}, see \cite{Capri:2015ixa,Capri:2016aqq,Capri:2018ijg} for details. As already reminded before, the topological partition function does not depend on the coupling $g$. This can also be shown using a BRST cohomological argument, as we reiterate in the next subsection. This means all observables can be computed in the $g\to0$ limit. Expanding around a nontrivial instanton background rather than around $A=0$  would lead to corrections of the type $e^{-1/g^2}$ into the effective action, but the latter vanish exponentially fast once $g\to0$ is considered. As such, we can a priori work around $A=0$.

This is good news, as explicit instanton computations are usually performed in a background gauge setting, being virtually impossible in other gauges such as Landau gauge. The above reasoning prevents us that we should resort to another gauge, such as the background Landau gauge, for which the Gribov problem and resolution is a bit different and actually far more complicated, in particular when BRST invariance is to be preserved \cite{Kroff:2018ncl,Dudal:2017jfw}. In \cite{Dudal:2017jfw}, such computation was presented for a constant temporal background, already complicated enough. For an $x$-dependent instanton background, the methodology of \cite{Dudal:2017jfw} simply looks technically impossible.

\subsection{Further evidence}

Before ending, we find it instructive to present yet another argument why a null Gribov parameter is also in full accordance with the possibility of a vanishing $\beta$-function discussed in \cite{Junqueira:2018zxr}. Indeed, the variation of the full action with respect to the coupling constant gives a BRST-exact term (up to boundary terms),
\begin{eqnarray} \label{deltaS}
\delta_g S = s \left( \Delta^{(-1)} \right)\;,
\end{eqnarray}
where $\Delta^{(-1)}$ is a polynomial in the fields and parameters, with ghost number equal to minus one. This result is independent of the gauge choice. Since the expectation values of BRST exact terms vanish, \eqref{deltaS} implies that
\begin{eqnarray}
\delta_g Z = \left\langle s \left( \Delta^{(-1)} \right) \right\rangle = 0\;,
\end{eqnarray}
with  $Z$ the generating functional. It means that the theory is insensitive to changes of the coupling constant, in other words, that the theory has no scale. This can be re-expressed by the theory not having a $\beta$-function, which makes impossible the feature of dimensional transmutation. Indeed, the Gribov gap equation is nothing but a tool giving $\gamma^2\propto \Lambda^2$, $\Lambda\sim \mu e^{-\frac{1}{\beta_0 g^2(\mu)}}$ being the fundamental scale of the theory if $\mu$ is the renormalization group scale; a quantity directly related to the $\beta$-function \cite{Celmaster:1979km}. However, in the absence of the latter, it holds that $\Lambda\equiv0$ and a classically massless (or better said scale invariant) theory will remain so at the quantum level. A rather similar situation showed up in the super $\mathcal{N}=4$ Yang-Mills theory which possesses a vanishing $\beta$-function. The absence of a renormalization group invariant scale makes it impossible to attach a dynamical meaning to the Gribov parameter, in such a way that the restriction to the first Gribov region is not required \cite{Capri:2014tta}.

\section*{Acknowledgments}

The work of ADP was partially supported by the DFG through the grant Ei/1037-1. The authors are grateful to the Conselho Nacional de Desenvolvimento Cient\'ifico e Tecnol\'ogico. The Coordena\c c\~ao de Aperfei\c coamento de Pessoal de N\'ivel Superior (CAPES) is acknowledged for financial support.

\appendix

\section{Absence of radiative corrections in topological Yang-Mills theories in presence of the Gribov parameter}

The proof that topological Yang-Mills theories in the presence of the Gribov-Zwanziger horizon function remain tree-level exact is outlined. Along the lines of \cite{Junqueira:2018xgl}, we need the tree-level propagators in order to show that all $n$-point functions are tree-level exact. The non-vanishing tree-level propagators which are relevant for the present work are computed from \eqref{intro10}. The results are given by
\begin{eqnarray}
\langle U^{ab}_\mu(-k) U^{cd}_\nu(k) \rangle &=&-\frac{1}{k^2}\delta_{\mu\nu}\delta^{abcd}\;,\nonumber\\
\langle V^{ab}_\mu(-k) V^{cd}_\nu(k) \rangle &=&-\frac{1}{k^2}\delta_{\mu\nu}\delta^{abcd}\;,\nonumber\\
\langle b^a(-k)b^b(k) \rangle&=&-2Ng^2\gamma^4\frac{1}{k^4}\delta^{ab}\;,\nonumber\\
\langle B^a_{\mu\nu}(-k) B^b_{\alpha\beta}(k) \rangle&=&-Ng^2\gamma^4\frac{1}{k^4}\delta_{\mu\nu\alpha\beta}\delta^{ab}\;,\nonumber\\
\langle A^a_\mu(-k) b^b(k) \rangle&=&-i\frac{k_\mu}{k^2}\delta^{ab}\;,\nonumber\\
\langle A^a_\mu(-k) B^b_{\alpha\beta}(k) \rangle&=&i\frac{1}{k^2}\Sigma_{\mu\alpha\beta}\delta^{ab}\;,\nonumber\\
\langle b^a(-k) U^{bc}_\mu(k) \rangle&=&i\sqrt{2}\gamma^2\frac{k_\mu}{k^4}f^{abc}\;,\nonumber\\
\langle B^a_{\mu\nu}(-k) U^{bc}_\alpha(k) \rangle&=&i\sqrt{2}g\gamma^2\frac{1}{k^4}\Sigma_{\alpha\mu\nu}f^{abc}\;,\label{prop}
\end{eqnarray}
while the vanishing tree-level propagators are
\begin{eqnarray}
\langle A^a_\mu(-k) A^b_\nu(k) \rangle = \langle A^a_\mu(-k) U^{bc}_\nu(k) \rangle = \langle b^a(-k) B^b_{\mu\nu}(k) \rangle &=& 0\;,\nonumber\\
\langle V^{ab}_\mu(-k) A^c_\nu(k) \rangle = \langle V^{ab}_\mu(-k) U^{cd}_\nu(k) \rangle = \langle V^{ab}_\mu(-k) B^c_{\alpha\beta}(k) \rangle = \langle V^{ab}_\mu(-k) b^c(k) \rangle &=& 0\;,\label{prop2}
\end{eqnarray}
with
\begin{eqnarray}
\varphi^{ab}_\mu&=&\frac{\sqrt{2}}{2}\left(U+iV\right)^{ab}_\mu\;,\nonumber\\
\bar{\varphi}^{ab}_\mu&=&\frac{\sqrt{2}}{2}\left(U-iV\right)^{ab}_\mu\;,\label{realdecomp}
\end{eqnarray}
and $U$ and $V$ being real fields. Moreover,
\begin{eqnarray}
    \Sigma_{\alpha\mu\nu}&=&\frac{1}{2}\left(\delta_{\alpha\mu}k_\nu-\delta_{\alpha\nu}k_\mu\right)\;,\nonumber\\
    \delta^{abcd}&=&\frac{1}{2}\left(\delta^{ac}\delta^{bd}-\delta^{ad}\delta^{bc}\right)\;,\nonumber\\
    \delta_{\mu\nu\alpha\beta}&=&\frac{1}{2}\left(\delta_{\mu\alpha}\delta_{\nu\beta}-\delta_{\mu\beta}\delta_{\nu\alpha}\right)\;.\label{proj}
\end{eqnarray}
The remaining propagators can be found in \cite{Junqueira:2017zea,Junqueira:2018xgl}. Hence, if we compare the present situation with the scenario of \cite{Junqueira:2018xgl}, we have the extra non-vanishing propagators given by \eqref{prop} together with four new vertices (see the local action \eqref{intro10}), namely: (i) $\bar{\varphi} A \varphi$, (ii) $\bar{\omega} A \omega$, (iii) $\bar{\omega} \varphi c$, and (iv) $\bar{\omega} A \varphi c$. Again, there is no vertex with $b$, so we cannot use $\langle bb \rangle$ to propagate $b$ to a loop diagram. Using the propagator $\langle BB \rangle$, we can only propagate an external $B$ to the vertex $BAA$, increasing the number of $A$ fields. This is the same cascade effect that occurs with the $\langle A B \rangle$ propagator as in \cite{Junqueira:2018xgl}. The new vertices (i), (ii) and (iii) have one $A$-leg. To not produce an internal $A$-leg we need to propagate it to an external field, but again $A$ only propagates through $\langle A B \rangle$ and $\langle A b \rangle$, producing only $B$ and $b$ as external legs, since the propagators with $A$ and the new fields vanish: $\langle A \omega \rangle = \langle A \bar{\omega} \rangle = \langle A \varphi \rangle = \langle A \bar{\varphi} \rangle = 0$.

The only possible problematic vertex is (iii), which does not possess $A$-legs, but we cannot propagate a vertex (iii) to another vertex (iii) because $\bar{\omega}$ only propagates to the vertex (ii) through $\langle \bar{\omega} \omega \rangle$; $c$ only propagates to the vertex $\bar{c} A c$ through $\langle \bar{c} c \rangle$; and $\varphi$ only to vertex (i) through $\langle \bar{\varphi} \varphi \rangle$, or to external legs $B$ and $b$ through $\langle \varphi B \rangle$ and $\langle \varphi b \rangle$, or to the vertex $BAA$ through $\langle \varphi B \rangle$. In the end, we can only propagate the vertex (iii) to vertices with internal $A$-legs or to external legs  $B$ and $b$. We conclude that all loop diagrams vanish, because we fall back to the same situation in which we can only construct a loop diagram with $B$ and $b$ as external legs, in order to avoid internal $A$-legs, but $\langle B \cdots B b \cdots b \rangle = \langle s(\text{something}) \rangle = 0$, due to BRST cohomology. Otherwise, it is impossible to close non-vanishing loops as we need gauge propagators to do it, and $\langle AA \rangle$ also vanishes in the presence of the local Gribov terms \eqref{AA=0}.

\section{On-shell gauge fixing }
The starting action is \eqref{gfaction} while the corresponding generating functional reads
\begin{equation}
Z_{gf}=\int \mathcal{D}\Phi e^{-S_{gf}}\;,\label{Zgf}
\end{equation}
where $\mathcal{D}\Phi=\mathcal{D}A\mathcal{D}\bar{\chi}\mathcal{D}\psi \mathcal{D}\bar{c}\mathcal{D}c\mathcal{D}\bar{\phi}\mathcal{D}\phi \mathcal{D}\bar{\eta}\mathcal{D}b\mathcal{D}B$. Our aim is to show that \eqref{Zgf} is equivalent to
\begin{equation}
Z_{FP}=\int \mathcal{D}A\mathcal{D}\psi\det(\Theta_\pm)\delta(\partial A)\delta(F_\pm)\delta(\partial\psi)\;,\label{Zfp}
\end{equation}
where $F_\pm=F\pm\tilde{F}$ and $\Theta_\pm$ as defined before in \eqref{thetadef}. We suppressed indices in the notation here.

Integration over the auxiliary fields $b$ and $B$ leads to
\begin{eqnarray}
Z_{gf}&=&\int [\mathcal{D}\Phi\mathcal{D}b\mathcal{D}B]\exp\left\{-\int d^4x\left[-\frac{1}{2\alpha}(\partial A)^2-\frac{1}{4\beta}F_\pm^2\right]-\int d^4x\left[
\left(\bar{\eta}^a-\bar{c}^a\right)\partial_\mu\psi^a_\mu+\bar{c}^a\partial_\mu D_\mu^{ab}c^b\right.\right.\nonumber\\
&-&\left.\left.\frac{1}{2}gf^{abc}\bar{\chi}^a_{\mu\nu}c^b\left(F_{\mu\nu}^c\pm\widetilde{F}_{\mu\nu}^c\right)-\bar{\chi}^a_{\mu\nu}\left(\delta_{\mu\alpha}\delta_{\nu\beta}\pm\frac{1}{2}\epsilon_{\mu\nu\alpha\beta}\right)D_\alpha^{ab}\psi_\beta^b+\bar{\phi}^a\partial_\mu D_\mu^{ab}\phi^b\right.\right.\nonumber\\
&+&\left.\left.gf^{abc}\bar{\phi}^a\partial_\mu\left(c^b\psi^c_\mu\right)\right]\right\}\;.\label{Zfp1}
\end{eqnarray}
Some inconvenient terms can be eliminated by the following shifts,
\begin{eqnarray}
\bar{\eta}^a&\longmapsto&\bar{\eta}^a+\bar{c}^a\;,\nonumber\\
\phi^b&\longmapsto&\phi^b-gf^{cde}(\partial_\nu D_\nu^{bc})^{-1}\partial_\mu\left(c^d\psi^e_\mu\right)\;,\nonumber\\
\bar{c}^a&\longmapsto&\bar{c}^a-\frac{1}{2}gf^{cde}\bar{\chi}^d_{\mu\nu}(F_\pm)^e_{\mu\nu}(\partial_\nu D_\nu^{ca})^{-1}
\end{eqnarray}
These transformations are valid perturbatively since $-\partial D>0$ \cite{Junqueira:2017zea,Junqueira:2018xgl}. Notice also that these shifts generate a trivial Jacobian. Hence,
\begin{eqnarray}
Z_{gf}&=&\int [\mathcal{D}\Phi\mathcal{D}b\mathcal{D}B]\exp\left\{-\int d^4x\left[-\frac{1}{2\alpha}(\partial A)^2-\frac{1}{4\beta}F_\pm^2\right]-\int d^4x\left[
\bar{\eta}^a\partial_\mu\psi^a_\mu+\bar{c}^a\partial_\mu D_\mu^{ab}c^b\right.\right.\nonumber\\
&-&\left.\left.\bar{\chi}^a_{\mu\nu}\left(\delta_{\mu\alpha}\delta_{\nu\beta}\pm\frac{1}{2}\epsilon_{\mu\nu\alpha\beta}\right)D_\alpha^{ab}\psi_\beta^b+\bar{\phi}^a\partial_\mu D_\mu^{ab}\phi^b\right]\right\}\;.\label{Zfp2}
\end{eqnarray}
Integration over the Faddeev-Popov and bosonic ghosts and the corresponding anti-ghosts leads to cancelling contributions,
\begin{eqnarray}
Z_{gf}&=&\int\;\mathcal{D}A\mathcal{D}\bar{\eta}\mathcal{D}\bar{\chi}\mathcal{D}\psi\;\exp\left\{-\int d^4x\left[-\frac{1}{2\alpha}(\partial A)^2-\frac{1}{4\beta}F_\pm^2\right]-\int d^4x\left[
\bar{\eta}^a\partial_\mu\psi^a_\mu\right.\right.\nonumber\\
&-&\left.\left.\bar{\chi}^a_{\mu\nu}\left(\delta_{\mu\alpha}\delta_{\nu\beta}\pm\frac{1}{2}\epsilon_{\mu\nu\alpha\beta}\right)D_\alpha^{ab}\psi_\beta^b\right]\right\}\;.\label{Zfp3}
\end{eqnarray}
Integration over $\bar{\eta}$ and $\bar{\chi}$ (see for instance \cite{Das:2019jmz}) subsequently leads to
\begin{equation}
Z_{gf}=\int\;\mathcal{D}A\mathcal{D}\psi\;\exp\left\{-\int d^4x\left[-\frac{1}{2\alpha}(\partial A)^2-\frac{1}{4\beta}F_\pm^2\right]\right\}\delta(\partial\psi)\delta(\Theta_\pm\psi)\;.\label{Zfp4}
\end{equation}
since $\bar{\chi}^a_{\mu\nu}$ is anti-symmetric, deriving w.r.t.~it will automatically anti-symmetrize the operator coupled to it, leading to the earlier introduced operator $\Theta_\pm$.

From the usual manipulations, the $\alpha$-term reproduces the usual delta for $\partial A$ and the $\beta$-term a delta for $F_\pm$,
\begin{equation}
Z_{gf}=\int\;\mathcal{D}A\mathcal{D}\psi\;\delta(\partial A)\delta(F_\pm)\delta(\partial\psi)\delta(\Theta_\pm\psi)\;.\label{Zfp4a}
\end{equation}
This expression shows that the gauge is fixed as we intended. Particularly, the extra constraint, $\delta(\Theta_\pm\psi)$ selects the correct physical spectrum, i.e.~the instanton modes.

An alternative and perhaps more insightful computation can be performed as follows. The field $\bar{\chi}$ is anti-symmetric and (anti-)self-dual. So we can fully anti-symmetrize it. Moreover, we can already use the other constraint $\delta(\partial \psi)$ to replace $\psi$ with $\psi^T$ (transverse) in the term $\bar{\chi} \ldots \psi^T$. The field $\bar{\chi}$, as an (anti-)self-dual tensor field, contains 3 degrees of freedom, just as the $\psi^T$. Let us denote it by $\bar{\chi}_{ind}$. Hence, we can say we have six Grassmann independent variables, which allows to schematically rewrite (see for instance \cite{Brooks:1988jm}) $\bar{\chi} D_\pm \psi^T \equiv  (\bar{\chi}_{ind}, \psi^T) * M * (\bar{\chi}_{ind},\psi^T)$. This matrix operator $M$ is six-dimensional and so is the Grassmann vector $(\bar{\chi}_{ind},\psi^T)$. Eventually, integration over the six-dimensional Grassmann vector leads to
\begin{equation}
Z_{gf}=\int\;\mathcal{D}A\mathcal{D}\psi\;\delta(\partial A)\delta(F_\pm)\delta(\partial\psi)\mathrm{Pfaff}(M)\;,\label{Zfp5}
\end{equation}
where $\mathrm{Pfaff}$ stands for the Pfaffian. From the general relation $\mathrm{Pfaff}(M) = \det^{1/2}(M) = \det\Theta_\pm$, we finally get
\begin{equation}
Z_{gf}=\int\;\mathcal{D}A\mathcal{D}\psi\;\delta(\partial A)\delta(F_\pm)\delta(\partial\psi)\det(\Theta_\pm)\;.\label{Zfp5}
\end{equation}
Of course, it is possible to cross from \eqref{Zfp4a} to \eqref{Zfp5} by evaluating the last $\delta$-constraint, keeping in mind the other constraints and the calculus rules to deal with Grassmann Jacobians \cite{Das:2019jmz}.

\bibliographystyle{utphys2}
\bibliography{library}

\end{document}